\documentclass[pmlr]{jmlr}

\usepackage{longtable}

\usepackage{booktabs}
\usepackage[load-configurations=version-1]{siunitx} 
\usepackage{amsmath}
\newcommand\norm[1]{\left\lVert#1\right\rVert}

\theorembodyfont{\upshape}
\theoremheaderfont{\scshape}
\theorempostheader{:}
\theoremsep{\newline}

\jmlrvolume{XXX}
\jmlryear{2019}
\jmlrworkshop{IJCAI Affective Computing Workshop}

\title[Emotion Recognition]{Learning Discriminative Features using Center Loss and Reconstruction as Regularizer for Speech Emotion Recognition}

  \author{\Name{Suraj Tripathi} \Email{surajtripathi93@gmail.com}\\
   \Name{Abhiram Ramesh} \Email{a.ramesh@samsung.com}\\
   \Name{Abhay Kumar} \Email{abhay1.kumar@samsung.com}\\
   \Name{Chirag Singh} \Email{c.singh@samsung.com}\\
   \Name{Promod Yenigalla} \Email{Promod.y@samsung.com}\\
   \addr Natural Language Understanding Group - Samsung Research, India}

\editor{Editor's name}

\begin{document}

\maketitle

\begin{abstract}
This paper proposes a Convolutional Neural Network (CNN) inspired by Multitask Learning (MTL) and based on speech features trained under the joint supervision of softmax loss and center loss, a powerful metric learning strategy, for the recognition of emotion in speech. Speech features such as Spectrograms and Mel-frequency Cepstral Coefficients (MFCCs) help retain emotion related low-level characteristics in speech. We experimented with several Deep Neural Network (DNN) architectures that take in speech features as input and trained them under both softmax and center loss, which resulted in highly discriminative features ideal for Speech Emotion Recognition (SER). Our networks also employ a regularizing effect by simultaneously performing the auxiliary task of reconstructing the input speech features. This sharing of representations among related tasks enables our network to better generalize the original task of SER. Some of our proposed networks contain far fewer parameters when compared to state-of-the-art architectures. We used the University of Southern California’s Interactive Emotional Motion Capture (USC-IEMOCAP) database in this work. Our best performing model achieves a 3.1\% improvement in overall accuracy and a 5.3\% improvement in class accuracy when compared to existing state-of-the-art methods.
\end{abstract}

\section{Introduction}
\label{sec:intro}
    Standard procedure in a majority of Natural Language Processing (NLP) solutions such as voice activated systems, chatbots etc., is to first convert speech input to text using Automatic Speech Recognition (ASR) systems and then apply NLP techniques to understand the said text \cite{seide2011conversational}. Human-computer interaction gets better as computers improve in predicting the current emotional state of the human speaker \cite{Imrieinproceedings}, in order to be capable of distinguishing between different contextual meanings of the same word. ASR resolves variations in speech from different users using probabilistic acoustic and language models \cite{hinton2012deep}, which results in speech transcriptions being speaker independent. This might be good enough for most applications, but is an undesired result for systems which rely on knowing the intended emotion to function correctly. \par
    Since the last decade, Deep Learning techniques have contributed significant breakthroughs in Natural Language Understanding (NLU). Deep Belief Networks (DBN) for SER, proposed by \cite{kim2013deep} and \cite{zheng2014eeg}, showed a significant improvement over baseline models \cite{Jin2015SpeechER, verarticle, mao2009multi, ntalampiras2011modeling, 4218125, Neiberg2006EmotionRI,Wu2011EmotionRO} that do not employ deep learning, which suggests that high-order non-linear relationships are better equipped for emotion recognition. \cite{chernykh2017emotion} trained DNNs on a sequence of acoustic features calculated over small speech intervals along with a probabilistic-natured CTC Loss function, which allowed the consideration of long utterances containing both emotional and unemotional parts and improved recognition accuracies. \cite{lee2015high} used a Bi-directional LSTM model to train feature sequences and achieved an emotion recognition accuracy of 62.8\% on the IEMOCAP \cite{busso2008iemocap} dataset, which is a significant improvement over DNN-ELM \cite{han2014speech}. \cite{satt2017efficient} used deep CNNs in combination with LSTMs to achieve better results on the IEMOCAP dataset. Our previous paper \cite{Yenigalla2018SpeechER} substituted speech transcriptions with phoneme sequences and spectrograms, retaining more of the emotion content in speech and beating the previous state-of-the-art results.\par
    \cite{yadav2018learning} proposed very deep VGG CNNs trained under the joint supervision of softmax and center loss to obtain highly discriminative deep features and deliver highly in inter-class segregation and intra-class compactness, ideal for Speaker Identification and Verification tasks. \cite{liebel2018auxiliary, ruder2017overview} have employed the powerful regularization technique of MTL in speech, language and vision related tasks to provide the network with an inductive bias, leading to a preference for model hypotheses that generalizes across different tasks. The IEMOCAP \cite{busso2008iemocap} experimental dataset used in this work suffers from class imbalance. The emotion Neutral covers almost half the data set, and the number of training samples per class is also restricted. Training a model with such data requires it to not only learn highly discriminative deep features from a limited dataset but also be general enough to identify emotion in new unheard speech inputs, as otherwise the models tend to overfit. To overcome these problems, in this paper, we have extended our work from \cite{Yenigalla2018SpeechER} to use parallel CNNs with center loss, regularized by an auxiliary task. As deep networks tend to overfit \cite{srivastava2014dropout} we have employed the use of parallel CNNs, thus extending the network in width without increasing computational costs. While we also considered contrastive loss \cite{chen2011extracting} and triplet loss \cite{7953194} for this work, both result in a substantial increase in data size as they group inputs into pairs and triplets respectively, whereas center loss adds no significant overhead to the training process. We have introduced an autoencoder inspired auxiliary task into our work with an aim to learn a representation as close as possible to the input speech feature. The idea here is that if the network simultaneously performs more than one task it will need to learn a representation that captures all tasks, thus reducing the odds of overfitting on the original task of SER.

\section{Proposed Methods}
    In this paper we consider speech features such as spectrogram and MFCC, which provide a deep neural network the necessary low level features required to accurately distinguish among different emotions. Experiments have been performed to show the effectiveness of center loss in generating intra-class compactness. Steps such as widening the network with the use of parallel CNNs and performing an auxiliary task to regularize the training process have also been taken to mitigate overfitting and achieve accuracies greater than existing state-of-the-art methods.
    \subsection{Model Architecture}
    2D filters in CNNs help capture 2D feature maps in a given input. Spectrogram and MFCC, which are representations of speech over time and frequency, contain additional information not available in just text, thus giving us further capabilities in our attempts to improve emotion recognition. \par
   \begin{figure}[!tb]
      \centering
      {\includegraphics[width=0.7\textwidth]{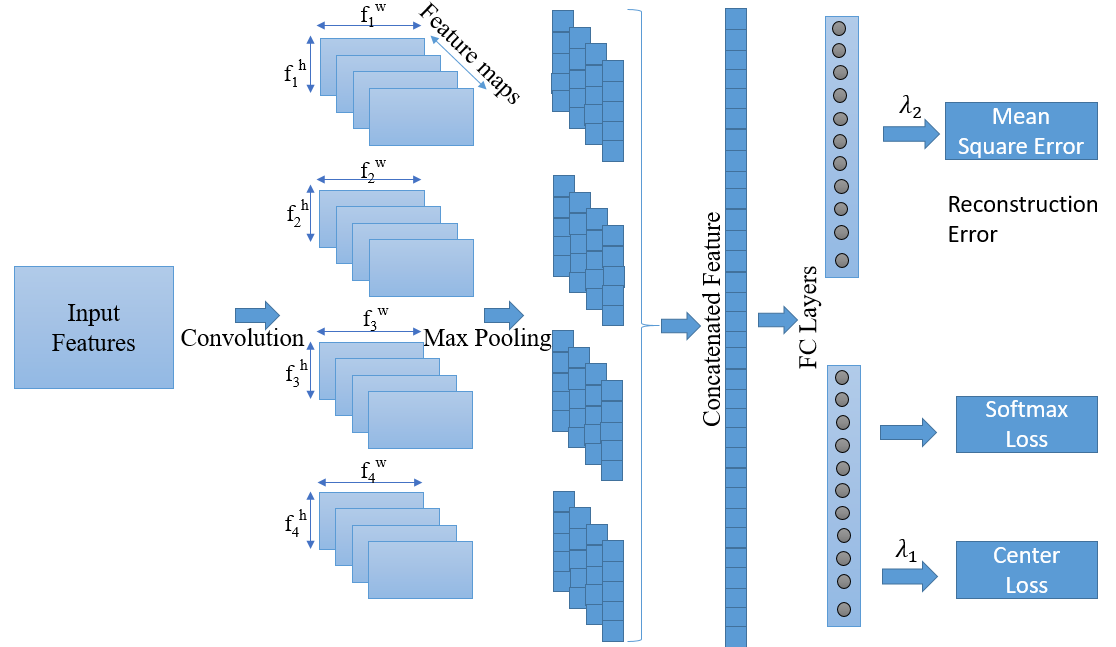}\label{fig:f1}}
      \caption{Model architecture}
\end{figure}
    Figure 1 details the shared 2D CNN architecture which takes in speech features as inputs. The architecture is an extension of our previous work on SER \cite{Yenigalla2018SpeechER}. 200 2D kernels are used for each of the 4 parallel convolution paths. These number of kernels and parallel CNN paths were observed to be optimal based on results on the validation data. Figuring out the optimal kernel size is a difficult and time taking task, which may depend on several factors all which cannot be clearly defined. To prevent choosing one single kernel size that could possibly be sub-optimal we decided to use kernels of different sizes, each of which is fixed for a single parallel path, to take advantage of the different patterns picked up by each kernel. The sizes of each of the 4 kernels in their respective parallel CNN paths are 4 x 6, 6 x 8, 8 x 10 and 10 x 12. These dimensions were also chosen after extensive experimentation and analysis of results on validation data. The features generated in the said convolution layers are then fed to their respective max-pool layers, which extracts 200 x 4 features from each parallel convolution path as the pool size is exactly half along the width and height of the convolution output. The extracted features are flattened, concatenated and then fed to two independent Fully Connected (FC) layers, which utilize Batch normalization. \par 
    We experimented with dropout rates varying between 25\% and 75\% for both the FC layers. The activation function used in the convolutional and FC layers is the Rectified Linear Unit (ReLU). These 2 FC layers are independently turned on and off based on the configuration of the network being trained, whereas the architecture prior to the FC layers are shared. The upper independent FC layer feeds its output to the Mean Square Error (MSE) loss (Decoder) layer when performing the auxiliary task, whereas the lower one feeds it to the softmax and center loss layers. The upper decoder layer consists of even more FC layers and a final output layer, which is dimensionally same as the input, producing the reconstructed speech features. The sigmoid activation function is applied to this output layer. MSE, relative to the input, is used here to calculate reconstruction loss. The softmax layer represents the output layer where each node represents one of the four emotion classes. Both softmax activation and cross-entropy loss functions are applied to this softmax layer. The center loss layer attempts to reduce intra-class separation. The different loss layers from which the gradients are back-propagated are configured based on the network. “Adadelta” optimization technique is used during training.
    \subsection{Feature Extraction}
    As the length of the utterances in the IEMOCAP dataset vary, we decided to limit its duration to 6 seconds (75th percentile of all audio lengths). Utterances shorter than 6 seconds are zero-padded. \par
    The presented models in this paper use spectrogram and MFCC as input to a 2D CNN. We used the "librosa" python package to compute the spectrogram and cepstral coefficients. As speech signals in the IEMOCAP corpus are sampled at 16 KHz, the sampling rate for spectrogram generation was also set to 16 KHz. Length of the “hann” and Fast Fourier Transform (FFT) windows is set to 2048, while the hop-length of the Short Term Fourier Transform (STFT) is set to 512. The Librosa python package then maps the obtained spectrogram magnitudes to the mel-scale to get mel-spectrograms. The hyper parameters and the python package used for MFCC generation are similar to the ones described above for spectrogram. These speech features are mel-scaled, putting emphasis on the lower end of the frequency spectrum over the higher ones, thus imitating the perceptual hearing capabilities of humans. 128 and 40 coefficients per window were generated for spectrogram and MFCC respectively. The input shapes of both spectrograms and MFCC were fixed on to after extensive experimentation, analysis of results on validation data and fine-tuning during the training process.
    \subsection{Joint Supervision}
    Center loss, which is widely used in face recognition tasks and known to minimize intra-class variations while maintaining separation between features of different classes, is used along with softmax loss during training. Center loss was preferred over its peers such as contrastive loss and triplet loss as it had the minimum overhead. Center loss is formulated as:
    
    \begin{equation}
    L_C = \frac{1}{2} \sum_{i=1}^n \norm{ f(x_i) - c_{y_i} }^2 
  \end{equation}
    \par
    where $f(x_i)$ denotes the deep feature extracted from the last hidden layer and $c_{y_i}$ $\in$ $R_d$ denotes the $y_{i}^{th}$ class center of deep features. As both softmax and center losses have jointly supervised the training process, the joint loss function is written as:
    \begin{equation}\label{second}
     L = L_S + \lambda_1 * L_C    
    \end{equation}

    \par
    where $L_S$ and $L_C$ represent softmax and center loss respectively and $\lambda_1$ represents the balance factor between the two. Different values in the range 0.001 to 5 were applied to $\lambda_1$, the best results being achieved at a value of 4. 
    
    \subsection{Reconstruction as Regularizer}
    When training a model for any task, the aim is to learn a good representation, which ideally is able to separate important information from data-dependent noise and also generalize well. Training for a single task increases the risk of overfitting to that task, whereas a model that trains on multiple related tasks simultaneously is able to learn better features by averaging the different noise patterns among all tasks. The secondary task in our work is inspired by an autoencoder and aims to reconstruct the input speech feature. It works towards minimizing the sum of the squared differences between the outputs of the logistic units and the input pixel intensities. This joint loss function, supervised by both softmax loss and the auxiliary task’s “reconstruction loss” can be represented as: 
    \begin{equation}\label{fourth}
     L = L_S + \lambda_2 * L_A    
    \end{equation}
    \par
    where $L_S$ and $L_A$ represents softmax loss and MSE respectively. The loss function for the model that performs the auxiliary task in addition to being jointly supervised by both softmax and center loss can be represented as:
    \begin{equation}\label{fifth}
     L = L_S + \lambda_1 * L_C + \lambda_2 * L_A   
    \end{equation}
    
    \par
    with both $\lambda_1$  and $\lambda_2$ being fine-tuned during the training process. The task of reconstructing the input speech feature was chosen to be the auxiliary task as, intuitively, the network would benefit from encoding important patterns present in the input speech feature, thus introducing an inductive bias, causing the model to capture a more general representation of our main SER task.
    
    \section{Dataset}
    We used USC’s IEMOCAP \cite{busso2008iemocap} database in this work. The IEMOCAP corpus comprises of five sessions where each session includes the conversation between two people, in both scripted and improvised topics and their corresponding labeled speech text. Each session is acted upon and voiced by both male and female voices and is without any speaker overlap among different sessions. Being consistent with prior research, only improvised data is used in this work as scripted text shows strong correlation with labeled emotions and can lead to lingual content learning, which can be an undesired side effect. While performing 5-fold cross validation, the data from 4 sessions is used for training in each fold. The data from the 5th session is split so as to use one speaker for validation while the other for accuracy testing. In our experiments, we have considered only 4 of them (Anger, Happiness, Sadness and Neutral) so as to remain consistent with earlier research. The final experimental dataset extracted from the original IEMOCAP data comprised of 4 classes named Neutral (48.8\% of the total dataset), Happiness (12.3\%), Sadness (26.9\%) and Anger (12\%). As there is data imbalance between different emotional classes we present our results on overall accuracy and average class accuracy.
    
\section{Discussion}
    In congruence with previous research efforts, we show the effectiveness of the proposed methods for emotion detection with our benchmark results on IEMOCAP dataset. In Table 1 we present our 5-fold cross-validation experimental results. Both overall and class accuracies are presented for better comparison, where overall accuracy is measured based on total counts irrespective of classes and class accuracy is the mean of accuracies achieved in each class. 
    
       \begin{table}[htbp]
\floatconts
  {tab:example}%
  {\caption{Comparison of model accuracies (*S = Softmax Loss, A = MSE (Auxiliary task), C = Center Loss)}}%
  {\begin{tabular}{llll}
  \bfseries Model & \bfseries Input & \bfseries Overall Accuracy &\bfseries Class Accuracy \\
  \cite{lee2015high} & Spectrogram & 62.8 & 63.9\\
  \cite{satt2017efficient} & Spectrogram & 68.8 & 59.4 \\
  \cite{Yenigalla2018SpeechER} & Spectrogram & 71.2 & 61.9 \\
  S & MFCC & 70.5 & 61.2 \\
  S+A & MFCC & 72.9 & 65.4 \\
  S+C & MFCC & 73.4 & 66.1\\
  S+A+C & MFCC & \textbf{74.1} & \textbf{66.7} \\
  S & Spectrogram & 71.2 & 61.9 \\
  S+A & Spectrogram & 73.1 & 65.2 \\
  S+C & Spectrogram & 73.6 & 66.5 \\
  S+A+C & Spectrogram & \textbf{74.3} & \textbf{67.2}
  \end{tabular}}
\end{table}
     Table 1 provides the results achieved on the SER task. We experimented with models that work with different combinations of softmax loss, center loss and auxiliary task based loss (MSE), which take in either spectrogram a or MFCC as input. Compared to the softmax only model we achieved better results with the addition of center loss and/or auxiliary task. We observed improvements in the S+A models as the introduction of the secondary task as a regularization method led to better SER accuracies. The same was observed with S+C, proving that targeting improvement in intra-class compactness led to improved discriminative abilities of deep features. Our best performing S+A+C models beat the state-of-the-art \cite{Yenigalla2018SpeechER} accuracies with the spectrogram based model outperforming by 3.1\% for overall accuracy and 5.3\% for class accuracy. L1 variation of equation (1) was also experimented with to achieve similar accuracies.
    To show the effectiveness of using center loss to create intra-class compaction, we have carried out a qualitative experiment where we visualize and compare 2D t-SNE \cite{maaten2008visualizing} projections of randomly selected 250 points from each of the 4 classes (Anger, Happiness, Sadness and Neutral) using the IEMOCAP dataset. These data points are extracted from the last hidden layers of the S only and S+C models for both spectrogram and MFCC. These embeddings were taken after training both the models for the same number of epochs. \par
    In Figure 2 and 3, we can see that models trained under softmax loss only had their embeddings scattered. Although fairly distinct structures can be observed, the separation between classes is not clear enough and some overlap can also be seen among the different classes except Happiness. On the other hand, the images representing the jointly supervised models (Models S+C for spectrogram and MFCC) show compact and distinct clusters proving the intuition behind using center loss to reduce intra-class separation. The number of parameters in our state-of-the-art beating S+C models is ~0.26 million compared to ~0.69 million in \cite{satt2017efficient} reducing the parameter size by ~62\%, thus also making our networks lighter with respect to memory requirements.

\begin{figure}[!tb]
\floatconts
{fig:example2}
{\caption{ (Spectrogram Input): Softmax only vs Softmax +
Center Loss}}
{%
\subfigure{%
\label{fig:Spectrogram, Softmax Loss}
\includegraphics[width=0.4\textwidth]{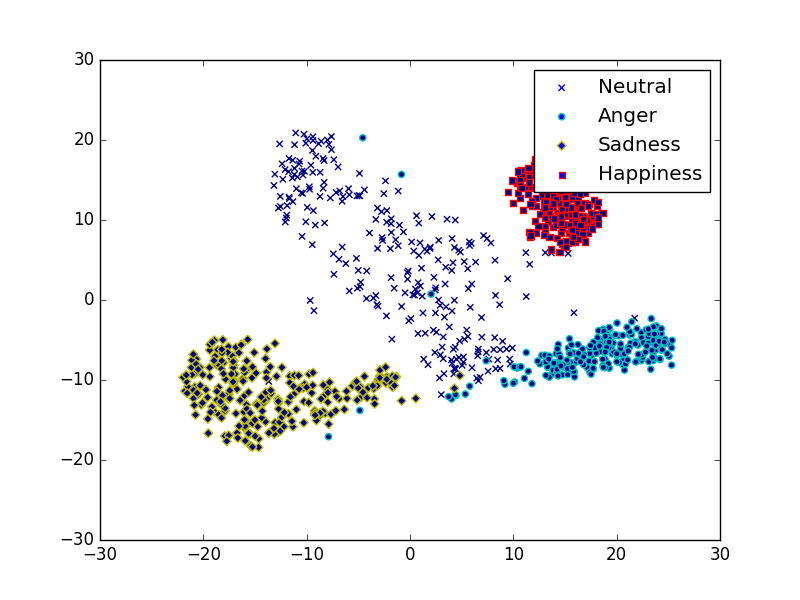}
}
\subfigure{%
\label{fig:Spectrogram, Softmax + Center Loss}
\includegraphics[width=0.4\textwidth]{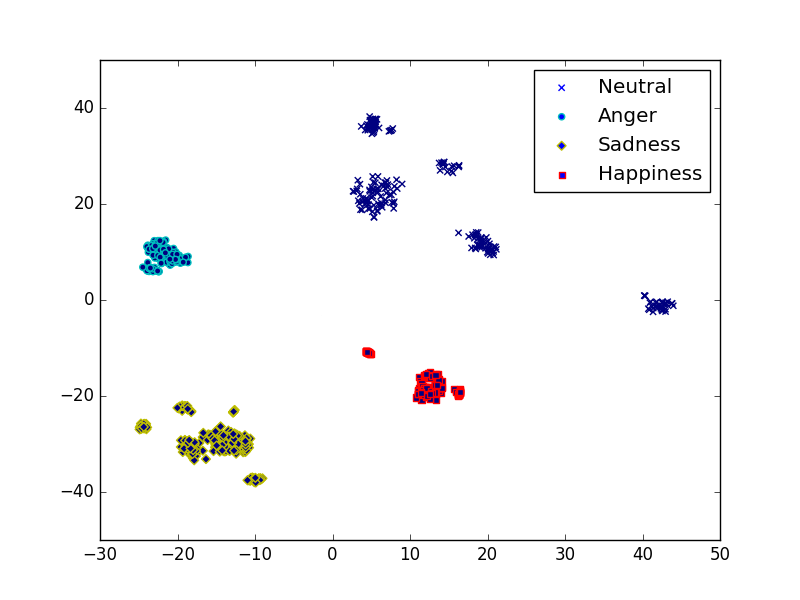}
}
}
\end{figure}

\begin{figure}[!tb]
\floatconts
{fig:example3}
{\caption{ (MFCC Input): Softmax only vs Softmax +
Center Loss}}
{%
\subfigure{%
\label{fig:Spectrogram, Softmax Loss}
\includegraphics[width=0.4\textwidth]{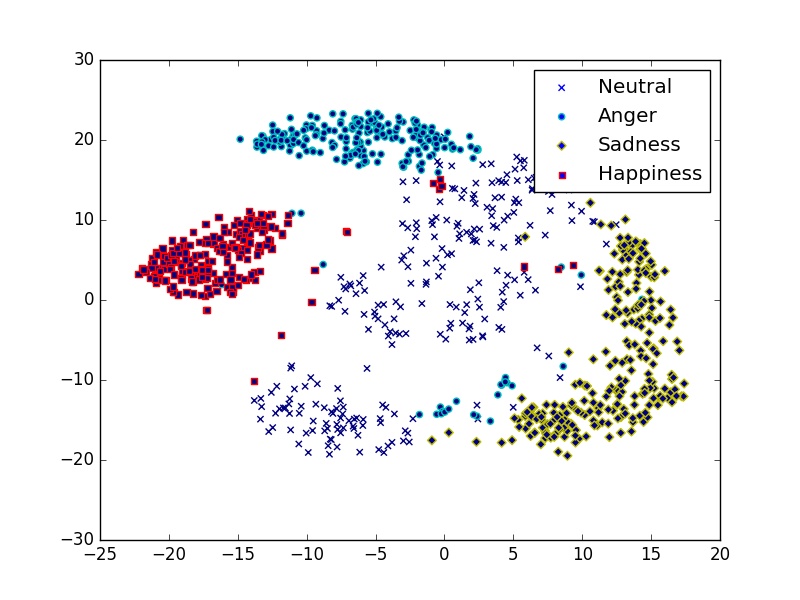}
}
\subfigure{%
\label{fig:Spectrogram, Softmax + Center Loss}
\includegraphics[width=0.4\textwidth]{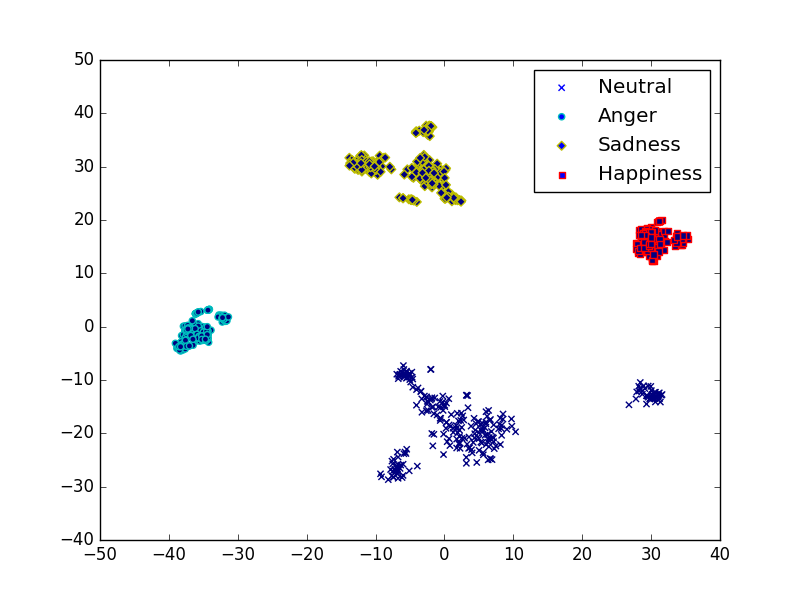}
}
}
\end{figure}

    \par
    As the data used in this work suffers from class imbalance and is also limited in size, it is important to make sure the model doesn’t overfit the training data. To encourage the model to learn a more generalized set of features, we have introduced the autoencoder inspired auxiliary task of reconstructing an as close as possible representation of the input speech feature. This secondary task is performed in addition to joint supervision by softmax and center loss. During training, this secondary task uses the input speech feature as the reconstruction target and attempts to minimize the MSE between the input to the CNN network and the output of the sigmoid layer. As the output of the sigmoid layer is always between 0 and 1 we scale down the values of the input speech feature to also lie between 0 and 1. This application of auxiliary task along with joint supervision by softmax and center loss helps achieve superior performance compared to S+C only models.
    
    \begin{table}[htbp]
\floatconts
  {tab:example}%
  {\caption{MSE on auxiliary task of input reconstruction with different model configurations (trained for same number of epochs)}}%
  {\begin{tabular}{lll}
  \bfseries Model & \bfseries Input & \bfseries MSE ($10^-4$)\\
  A & MFCC & 21.32 \\
  S+A & MFCC & 03.62 \\
  S+A+C & MFCC & \textbf{03.47} \\
  A & Spectrogram & 18.31 \\
  S+A & Spectrogram & 03.23 \\
  S+A+C & Spectrogram & \textbf{03.29}
  \end{tabular}}
\end{table}
    \par
    Table 2 shows the performance of the auxiliary task of reconstructing the input speech feature for different configurations of the network on the test set. Relative to the network which simply tries to reconstruct the input, A only, a clear improvement in loss can be seen in networks which also perform SER with softmax and center loss, suggesting that this joint supervision has also positively affected the performance of the auxiliary task.
    
    \section{Conclusion}
    In this paper we have proposed a CNN based architecture that works with speech features and is trained under the joint supervision of softmax loss and center loss, while also performing the auxiliary task of input speech feature reconstruction. We achieved greater inter-class separation and intra-class compactness by introducing center loss to the learning process, which to the best of our knowledge has not been used in SER tasks. The addition of an auxiliary task to the learning process helped regularize the model, generating even more general deep features and further improving SER accuracies. Our proposed S+C model is significantly smaller (~62\%) with respect to the number of parameters it contains when compared to state-of-the-art architecture in \cite{satt2017efficient}, while our S+A+C (spectrogram) model outperforms the benchmark results by over 3\% and 5\% for overall and class accuracies respectively. 
    When tested in live scenarios our model’s inference time was around 30 micro-seconds suggesting that the proposed model can be used for real-time emotion related applications such as conversational chat-bots, social robots etc., where identifying emotion and sentiment hidden in speech may play a role in better conversation.
\bibliography{sample.bib}
\end{document}